\documentclass[twocolumn,showpacs,preprintnumbers,amsmath,amssymb,prb]{revtex4}

\usepackage{graphicx}
\usepackage{dcolumn}
\usepackage{bm}

\begin{document}


\title{Surface and vortex structures 
in noncentrosymmetric superconductors 
under applied magnetic fields}

\author{Masatosi Oka}
\author{Masanori Ichioka}
\author{Kazushige Machida}
\affiliation{
Department of Physics, Okayama University, Okayama 700-8530, Japan
}

\date{\today}

\begin{abstract}
We investigate how the macroscopic spatial structure
of broken inversion symmetry manifests
in noncentrosymmetric superconductors,
by the microscopic broken inversion symmetry of the crystal structure.
Based on the time-dependent Ginzburg-Landau theory including
the Pauli paramagnetic effect and the Rashba interaction,
we demonstrate that the centrosymmetric structures
of the internal field and the screening current are broken macroscopically.
The flow structure of paramagnetic supercurrent
spontaneously induce the flux flow without applying external currents.
\end{abstract}

\maketitle

\section{INTRODUCTION}
Recently, several heavy fermion superconductors 
breaking inversion symmetry in the crystal structure 
were successively discovered, such as ${\rm CePt_3Si}$,\cite{Bauer} 
${\rm UIr}$,\cite{Akazawa} and ${\rm CeRhSi_3}$,\cite{Kimura}
and attract much attension. 
In the noncentrosymmetric superconductors,
there appear some exotic properties
through the spin-orbit coupling of the Rashba interaction
which induces Fermi surface splitting between two helical spin projections.\cite{Rashba,Sugawara}
For the superconductivity on the Fermi surfaces,
the special pairing form is only possible in the case of the spin-triplet pairing channel,
or the upper critical field can exceed the Pauli-Clogston limiting field
even for the spin-singlet pairing.\cite{Frigeri}
Furthermore, application of a magnetic field
makes the superconductivity more exotic,
such as helically modulated superconducting order parameter,\cite{Kaur}
and spin polarization
${\bf M}_{\rm s}\varpropto{\bf n}\times{\bf J}_{\rm s}$
induced by a supercurrent ${\bf J}_{\rm s}$,
where ${\bf n}$ is the unit vector along the polar axis of the noncentrosymmetry.
Then, the paramagnetic supercurrent $\nabla\times{\bf M}_{\rm s}$ also appears
in association with ${\bf M}_{\rm s}$.\cite{Edelstein,Yip,Fujimoto}

The purpose of this paper is to show how the macroscopic structures
of broken inversion symmetry appear in noncentrosymmetric superconductors
whose microscopic inversion symmetry is broken in the crystal structure,
with a view to detecting experimentally direct evidence of the noncentrosymmetric superconductors.
In the presence of the time reversal symmetry at a zero field,
we can not expect the broken-centrosymmetric spatial structure,
since the band structure in momentum space still conserves
the inversion symmetry even in the noncentrosymmetric crystal
structure.\cite{Samokhin}
Under magnetic fields,
by the combined two broken inversion symmetries of space and time,
exotic noncentrosymmetric spatial structures will be revealed.
We will see how these structures of the internal field and the screening current appear both in
the Meissner state and in the mixed state by the paramagnetic effect,
using the simulation of the time-dependent Ginzburg-Landau (TDGL) theory.

We describe the TDGL equation for
the noncentrosymmetric superconductors
and explain the method of our calculation in Sec. II.
In Sec. III, we consider noncentrosymmetric spatial structures
of the screening current and the penetrating internal magnetic field
near surfaces in the Meissner state.
In Sec. IV, we study exotic spatial structures of the current and
the internal field around a vortex. These structures induce the spontaneous flux flow
without applying external currents.
The last section is devoted to the summary.

\section{NUMERICAL SIMULATION FOR NONCENTROSYMMETRIC SUPERCONDUCTORS}
Numerical simulation of the TDGL equation is one of methods
to investigate dynamical and static properties of superconductivity.
The theory assumes that time evolutions of an order
parameter $\Delta$ and a vector potential ${\bf A}$ 
depend on the relaxation processes 
of a superconducting free energy to the equilibrium. 
Thus, the TDGL equation is obtained by \cite{Gorkov}
\begin {equation} 
\frac{{\partial}{\Delta}}{{\partial}t}
=-\frac{{\delta}F}{{\delta}{\Delta}^{\ast}}, 
\qquad
\frac{{\partial}{\bf A}}{{\partial}t}
=-\frac{{\delta}F}{{\delta}{\bf A}}, 
\label{eq:TDGL0}
\end{equation}
in an appropriate unit for times, 
where $F$ is a superconducting free energy, 
and ${\delta}/{{\delta}x}$ indicates a functional derivative
with respect to $x$. 

Hamiltonian of a noncentrosymmetric $s$-wave superconductor 
with the Rashba interaction is expressed as
\begin{eqnarray}
\mathcal{H}={\int}d{\bf r}\left\{\psi^{\dagger}
\left(\frac{p^2}{2m}\sigma_0+
\left(\alpha{\bf p}\times{\bf n}+\mu_B{\bf B}\right)
\cdot{\bf \sigma}\right)\psi\right.
\nonumber \\ 
+\left.
g\psi^{\dagger}i\sigma_y\psi{\psi}^{\dagger}i\sigma_y\psi\right\} , 
\label{hamiltonian} 
\end{eqnarray}
with the electron momentum ${\bf p}=-i\hbar\nabla$,
the spin-orbit coupling constant $\alpha$, 
the Bohr magneton $\mu_B$, the pairing interaction $g$,  
and the $2\times2$ unit matrix $\sigma_0$, and
the Pauli matrix $\sigma=(\sigma_x,\sigma_y,\sigma_z)$. 
$\psi$ is a field operator of the conduction electrons.  
For simplicity, 
we consider the case of a single component order parameter, 
neglecting the mixture of different parity components of 
the order parameter.\cite{Gorkov2,Frigeri} 
We also neglect the interband pairing between split bands of 
$\pm{\bf p}\times{\bf n}$ spin projections, since we consider 
the case when the band splitting energy due 
to the Rashba-interaction is enough large,  
as shown in the band structure calculation of 
${\rm CePt_3Si}$.\cite{Samokhin} 

From the Hamiltonian (\ref{hamiltonian}), the GL free energy 
is obtained as \cite{Kaur,Edelstein2} 
\begin{eqnarray} &&
F={\int}{\rm d}V\left[
|a| \left|\Psi\right|^2
+\frac{b}{2}\left|\Psi\right|^4 
\right. 
\nonumber \\ &&
+\gamma\left|\left(-i\hbar\nabla
  +\frac{2\pi}{\phi_0}{\bf A}\right)\Psi\right|^2
+\frac{1}{8\pi}\left|{\bf B}-{\bf H}\right|^2 
\nonumber \\ &&
+\varepsilon{\bf n}\cdot{\bf B}\times
\left.\left\{ \Psi^\ast\left(-i\hbar\nabla
+\frac{2\pi}{\phi_0}{\bf A}\right)\Psi+
\textit{c.c.} \right\}\right], 
\label{eq:F1}
\end{eqnarray}
for the order parameter $\Psi$, 
where $a$, $b$, and $\gamma$ are coefficients 
depending on the details of Fermi surface structure, 
$\phi_0$ is the flux quantum.  
${\bf B}$ and ${\bf H}$ are, respectively, 
internal and external magnetic fields.  
The last term with material dependent coefficient 
$\varepsilon$ $(\propto \alpha \mu_{\rm B})$ appears 
due to the absence of the inversion symmetry. 
The expression of Eq. (\ref{eq:F1}) is also applicable to 
the single component case 
in $p$-wave pairing, as well as $s$-wave pairing. 
In a dimensionless form, Eq. (\ref{eq:F1}) is written as
\begin{eqnarray}
F={\int}{\rm d}V\left\{-|\Delta|^2+\frac{1}{2}|\Delta|^4+\frac{1}{1-T}
\left|\left(-i\nabla+{\bf A}\right)\nabla\right|^2\right.
\nonumber\\
+\left.\frac{\kappa^2}{\left(1-T\right)^2}
\left(\left|{\bf B}-{\bf H}\right|^2
-\varepsilon{\bf n}\cdot{\bf B}\times{\bf J}_{\rm s}\right)\right\} ,
\label{eq:F2}
\end{eqnarray}
where  $\Delta$ is the normalized order parameter,
$\kappa$ is the GL parameter,
${\bf J}_{\rm s}$ is the supercurrent density defined by
\begin{eqnarray}
{\bf J}_{\rm s}\equiv-\left(1-T\right)\text{Re}\left[\Delta^\ast
\left(-i\nabla+{\bf A}\right)\Delta\right]/\kappa^2.
\label{eq:Js}
\end{eqnarray}
Here, $\varepsilon$ is normalized by $\pi a_0\xi_0^3/\hbar\phi_0$,
where $a_0$ and $\xi_0$ are zero temperature values of $a$ and 
coherence length, respectively. 
Using Eqs. (\ref{eq:TDGL0}) and (\ref{eq:F2}), 
we obtain the TDGL equation coupled with Maxwell
equations 
\begin{eqnarray}
\frac{\partial\Delta}{{\partial}t}=-\left(1-T\right)
\left(-\Delta+\Delta^2\Delta^\ast\right)
+\left(-i\nabla+{\bf A}\right)^2\Delta 
\nonumber \\
+\varepsilon\left({\bf n}\times{\bf B}\right)
\cdot\left(-i\nabla+{\bf A}\right)\Delta
+\frac{\varepsilon}{2}\Delta\left(-i\nabla\right)
  \cdot\left({\bf n}\times{\bf B}\right)  , 
\label{b}
\\ 
\frac{1}{\kappa^2}\frac{\partial{\bf A}}{{\partial}t}
=-\frac{1-T}{\kappa^2}\text{Re}
\left[\Delta^\ast\left(-i\nabla+{\bf A}
  +\varepsilon{\bf n}\times{\bf B}\right)\Delta\right]
\nonumber\\
-\nabla\times\left({\bf B}-{\bf M}_{\text {s}}\right) , 
\label{c}
\end{eqnarray}
where the spin polarization
${\bf M}_{\rm s} \propto {\bf n}\times{\bf J}_{\rm s} $
is defined as 
\begin{eqnarray}
{\bf M}_{\text {s}}
\equiv-\varepsilon\frac{1-T}{\kappa^2}{\bf n}\times\text{Re}
\left[\Delta^\ast\left(-i\nabla+{\bf A}\right)\Delta\right] . 
\label{d}
\end{eqnarray}
In the case of uniform ${\bf B}$, we see from Eq. (\ref{b})
that $\Delta$ has phase factor of the helical state as 
$\Delta({\bf r})=\tilde{\Delta}({\bf r}) 
{\rm e}^{-i\varepsilon({\bf n}\times{\bf B})\cdot{\bf r}}$.\cite{Kaur} 
When this form of $\Delta$ is substituted to Eq. (\ref{c}), 
the $\varepsilon$-dependence in  
$\Delta^\ast\left(-i\nabla+{\bf A}
  +\varepsilon{\bf n}\times{\bf B}\right)\Delta$ vanishes.
Then, no additional diamagnetic current are induced by
the phase factor of the helical state in the case of the uniform ${\bf B}$.
On the contrary, $\nabla\times{\bf M}_{\text {s}}$ in Eq. (\ref{c})  
gives the $\varepsilon$-dependence due to the phase of the helical state,  
because $\varepsilon{\bf n}\times{\bf B}$ is absent in Eq. (\ref{d}).
When ${\bf B}$ has spatial variation,
the phase factor of the helical state becomes nontrivial,
and the $\varepsilon$-dependence appears also from other terms
in Eqs. (\ref{b})-(\ref{d}).
Our calculations totally include
contributions from the phase factor of
the helical order parameter in Eq. (\ref{b}),
spin polarization ${\bf M}_s$,
and effects of nonuniform ${\bf B}$.
These induce the noncentrosymmetric structure in the macroscopic length scale.

\begin{figure}
\begin{center}
\resizebox{65mm}{70mm}{\includegraphics{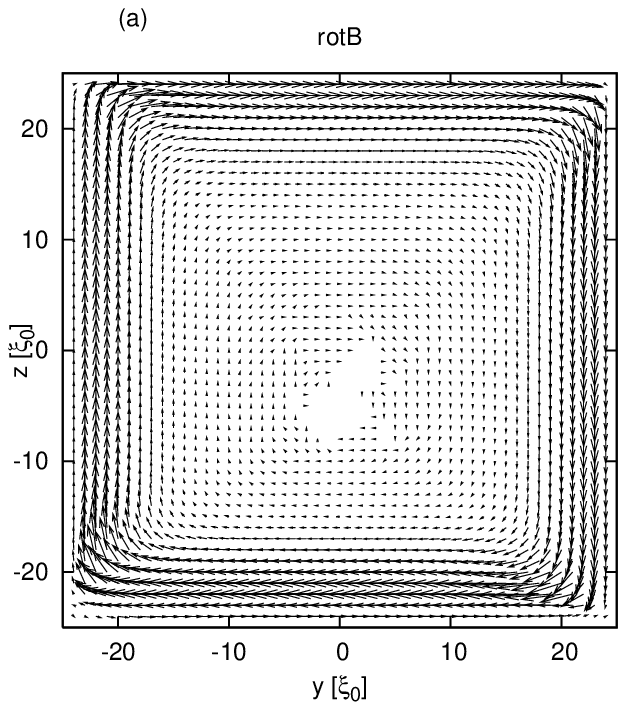}}
\resizebox{65mm}{70mm}{\includegraphics{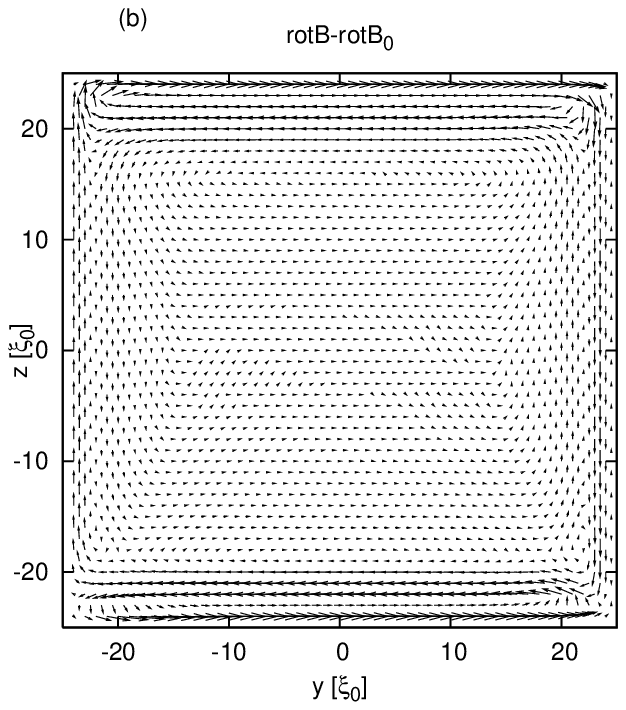}}
\caption{\label{cur_Meis} 
Spatial distribution of the current flow in the Meissner state.
We plot $\nabla\times{\bf B}$ in (a) and 
$\nabla\times{\bf B}-\nabla\times{\bf B}_0$ in (b) 
for the region of $50\xi_0 \times 50\xi_0$ with open boundary. 
The length of arrow indicates the amplitude of the current.
}
\end{center}
\end{figure}


\begin{figure}
\begin{center}
\resizebox{70mm}{50mm}{\includegraphics{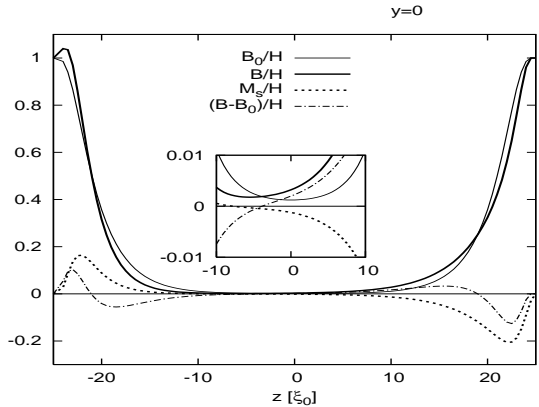}}
\caption{\label{intfld_Meis_sec} 
Profile of the internal magnetic field in the Meissner state 
along a plane parallel to ${\bf n}$ at $y=0$. 
We plot $B$, $B_0$, $B-B_0$ and $M_{\rm s}$ as a function of $z$. 
The inset focuses the behavior at the bulk region. 
}
\end{center}
\end{figure}

We perform numerical simulations following the time-evolution given by 
Eqs. (\ref{b})-(\ref{d}), as in previous works 
for centrosymmetric superconductors.\cite{KatoC,Kato,MMachida,Matsunaga}
We consider the case  ${\bf n}=\left(0,0,1\right)$ as in 
CePt$_{3}$Si and CeRhSi$_{3},$\cite{Yamamoto,Frigeri2}
and apply the magnetic field ${\bf H}=(H,0,0)$ along the $x$ axis. 
The calculations are done in a two dimensional square region  
parallel to  $yz$ plane. 
The region size is $50\xi_0\times50\xi_0$ 
(i.e., $-25 \le y/\xi_0 \le 25$ and $-25 \le z/\xi_0 \le 25$).  
Outside of the region we set $\Delta=0$ and ${\bf B}={\bf H}$. 
In our calculation, 
$T=0.5T_{c,H=0,\varepsilon=0}$ and $\kappa=2.0$, 
which is enough large $\kappa$ to study the qualitative properties 
in type-II superconductors, while we use smaller $\kappa$ than that of 
${\rm CePt_3Si}$ for the reason of the system size 
and the efficiency of the simulation.   
We set $\varepsilon=1$ and qualitatively study the contribution 
of the $\varepsilon$-dependent terms due to the noncentrosymmtry. 

\section{SURFACE STRUCTURE IN THE MEISSNER STATE}
First, we discuss the screening current and the penetrating internal 
field near surfaces in the Meissner state, when  
$H=0.04H_{c2,T=0,\varepsilon=0}$. 
We show the flow of the screening current $\nabla \times {\bf B}$ 
in Fig. \ref{cur_Meis} (a), and the profile of the internal field ${\bf B}$ 
along a plane parallel to {\bf n} at $y=0$ in Fig. \ref{intfld_Meis_sec}.
We see that, due to the screening current along the surface, 
$B$ is dumped toward inside of the superconductor. 
We also calculate the internal field ${\bf B}_0$ when $\varepsilon=0$ 
for the case of the conventional superconductor with centrosymmetry. 
Then, to estimate the contribution of the $\varepsilon$-dependent terms 
due to the noncentrosymmetry in GL Eqs. (\ref{b})-(\ref{d}), 
we show the difference $\nabla\times{\bf B}-\nabla\times{\bf B}_0$ 
in Fig. \ref{cur_Meis}(b) and $B-B_0$ in Fig. \ref{intfld_Meis_sec}.  
From this difference, we see that spatial structure of $B$ breaks 
centrosymmetry in the direction of ${\bf n}$. 
That is, ${\bf B}({\bf r})$ is not symmetric 
for the inversion $z \leftrightarrow -z$.  
Qualitatively, the noncentrosymmetric structure of ${\bf B}$ is due 
to the contribution of ${\bf M}_{\rm s}$
appearing near surfaces at $z=\pm 25 \xi_0$. 
We also show  $M_{\rm s}$ in  Fig. \ref{intfld_Meis_sec}.  
The contribution of $\varepsilon$ also appears
from other terms than ${\bf M}_{\rm s}$,
such as the phase factor of the helical state,
when ${\bf B}$ has spatial variation.
Near the surface, ${\bf M}_{\rm s}$ dominantly contributes to
the sign of ${\bf B}-{\bf B}_0$.
Far from the surface, there appear other contributions
compensating ${\bf M}_{\rm s}$.
As is shown in Fig. \ref{cur_Meis}(b), the paramagnetic supercurrent
$\nabla\times{\bf M}_{\rm s}$  
flows toward the same $+y$-direction near both surfaces 
at $z=\pm 25 \xi_0$.  
Therefore, near the surface at $z=25\xi_0$ ($z=-25\xi_0$),  
the screening current $\nabla\times{\bf B}$ is 
enhanced (suppressed) so that $B$ is suppressed (enhanced) 
compared with $B_0$.  
This indicates that the field penetration is different between 
$+{\bf n}$ and $-{\bf n}$-directions.\cite{Yip2} 

In the closed system as in our case,
due to the current conservation,
the backflow current to the paramagnetic current flows 
toward the opposite $-y$-direction far from the surface.   
Therefore, contraly to the region near the surface, 
the screening current $\nabla\times{\bf B}$ is 
suppressed (enhanced), so that $B$ is enhanced (suppressed) 
compared with $B_0$ in the inside region $-4\xi_0<z<19\xi_0$ 
($-21\xi_0<z<-4\xi_0$). 

On the other hand, in the profile of $B$ along a plane 
perpendicular to ${\bf n}$ at $z=0$, 
$B \sim B_0$, i.e., $B$ has symmetric profile for 
the inversion $y \leftrightarrow -y$. 

\section{VORTEX STRUCTURE IN THE MIXED STATE}
\begin{figure}
\begin{center}
\begin{tabular}{cc}
\resizebox{40mm}{47mm}{\includegraphics{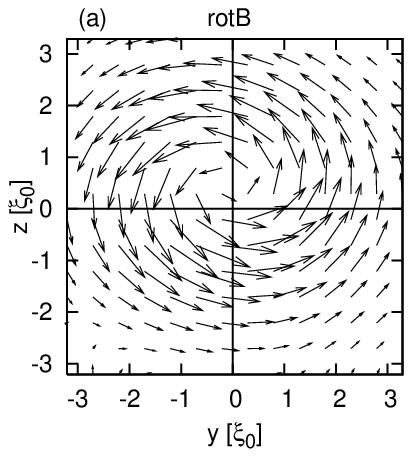}}&
\resizebox{40mm}{47mm}{\includegraphics{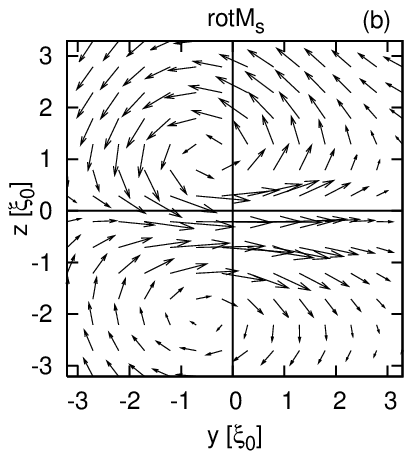}}\\
\end{tabular}
\caption{\label{cur_vor}
Spatial distribution of the current flow around a vortex.
We plot $\nabla\times{\bf B}$ in (a) and 
$\nabla\times{\bf M}_{\rm s}$ in (b) 
for a focused region of $6\xi_0 \times 6\xi_0$. 
The length of arrow indicates the amplitude of the current. 
The vortex center is located at $(y,z)=(0,0)$.
}
\end{center}
\end{figure}

\begin{figure}
\begin{center}
\begin{tabular}{cc}
\resizebox{40mm}{60mm}{\includegraphics{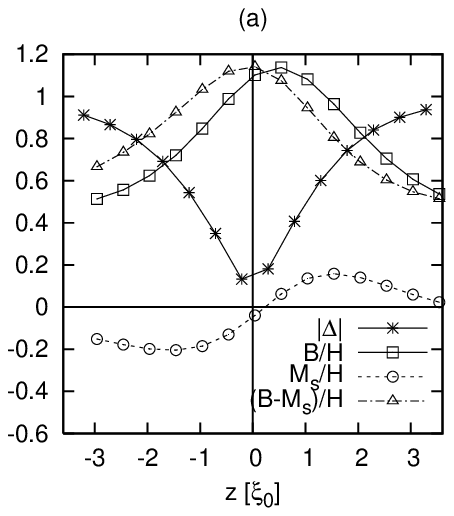}}&
\resizebox{40mm}{60mm}{\includegraphics{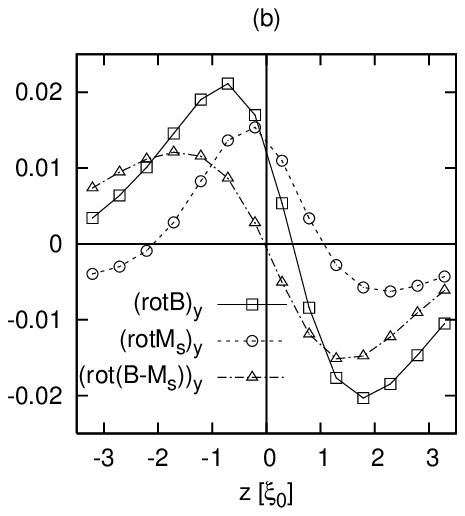}}\\
\end{tabular}
\caption{\label{vor_sec} 
Profile of the internal magnetic field (a) and the current (b) 
around a vortex along the plane parallel to ${\bf n}$
through the nearest meshpoint to the vortex center.
We plot $B$, $M_{\rm s}$, $B-M_{\rm s}$ and $|\Delta|$ in (a) and 
$(\nabla\times{\bf B})_y$, $(\nabla\times{\bf M}_{\rm s})_y$, 
$(\nabla\times{\bf B})_y-(\nabla\times{\bf M}_{\rm s})_y$ in (b). 
}
\end{center}
\end{figure}

Next, we discuss the vortex structure  
in the mixed state, when $H=0.15H_{c2,T=0,{\varepsilon}=0}$. 
In Fig. \ref{cur_vor}(a), we show the flow of 
the current $\nabla\times{\bf B}$, picking out $6\xi_0 \times 6\xi_0$ region 
around a vortex. 
While we set the position of the vortex center 
as $(y,z)=(0,0)$ in the figure,   
the winding center of the current does not coincide to the vortex center, 
shifting to the ${\bf n}$-direction.  
This is due to the paramagnetic supercurrent $\nabla\times{\bf M}_{\rm s}$. 
As shown in Fig. \ref{cur_vor}(b), $\nabla\times{\bf M}_{\rm s}$ flows toward 
the $+y$ direction at the vortex core. 
And backflow of $\nabla\times{\bf M}_{\rm s}$ 
returns to $-y$-direction outside of vortex.  
Figure \ref{vor_sec} shows the profile of the internal field 
and the current along the plane parallel to ${\bf n}$
through the nearest meshpoint to the vortex center. 
In Fig. \ref{vor_sec}(a), we also show  $M_{\rm s}$ and $B-M_{\rm s}$ 
in addition to $B$. 
The diamagnetic component $B-M_{\rm s}$ is almost symmetric for 
$z \leftrightarrow -z$, as in the conventional centrosymmetric superconductors. 
The peak position of $B-M_{\rm s}$ coincides to the vortex center.
However, $M_{\rm s}$ shows almost antisymmetric structure 
for $z \leftrightarrow -z$, since ${\bf J}_{\rm s}$ of 
${\bf M}_{\rm s} \propto {\bf n}\times{\bf J}_{\rm s}$ changes the flow direction at $z\sim0$.  
Summing up $B-M_{\rm s}$ and $M_{\rm s}$, we understand that $B$ 
has noncentrosymmetric structure at the vortex. 


\begin{figure}
\begin{center}
\resizebox{80mm}{!}{\includegraphics{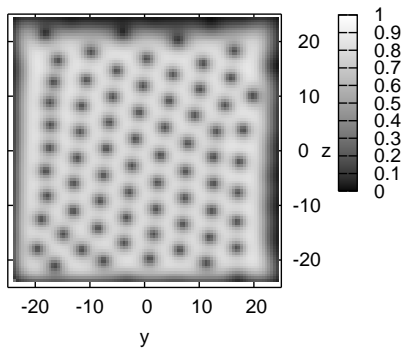}}
\caption{\label{vor_lat} 
Snapshot for the spatial structure of $|\Delta({\bf r})|$ 
in the mixed state without applying external currents.\cite{movie}
}
\end{center}
\end{figure}

In Fig. \ref{vor_sec}(b), we plot the $y$-component of the current 
$(\nabla\times{\bf B})_y$ with $(\nabla\times{\bf M}_{\rm s})_y$ 
and $(\nabla\times{\bf B})_y-(\nabla\times{\bf M}_{\rm s})_y$. 
The flow of the diamagnetic current component  
$(\nabla\times{\bf B})_y-(\nabla\times{\bf M}_{\rm s})_y$ 
is winding around the vortex center. 
The slope of $M_{\rm s}$ in Fig. \ref{vor_sec}(a) gives the 
paramagnetic current component $(\nabla\times{\bf M}_{\rm s})_y$. 
Since $M_{\rm s}$ is increasing near $z=0$ in Fig. \ref{vor_sec}(a), 
$(\nabla\times{\bf M}_{\rm s})_y$ flows $+y$ direction at the vortex core. 
Therefore, the position where $(\nabla\times{\bf B})_y=0$ 
deviates from $z=0$ to $+z$-direction. 
The noncentrosymmetric vortex structure, breaking cylindrical symmetry,
appears due to the antisymmetric spatial structure
of the spin polarization ${\bf M}_{\rm s}$.

On the other hand, in the profile along a plane
perpendicular to {\bf n} through the nearest meshpoint to the vortex center, 
$M_{\rm s} \sim 0$ and $(\nabla\times{\bf M}_{\rm s})_z \sim 0$, 
so that $B$ is symmetric and $(\nabla\times{\bf B})_z$ 
is antisymmetric at $z=0$ for the inversion $y \leftrightarrow -y$, 
as in the centrosymmetic superconductors. 

The flow structure of the paramagnetic supercurrent at the vortex gives similar effects
as in the situation when an external current is applied toward $+y$ direction.
Then, in our simulation, the paramagnetic
supercurrent spontaneously induces the flux flow even in the absence of the external current.
Figure \ref{vor_lat} shows the snapshot of the spatial structure 
of $|\Delta({\bf r})|$ when $H=0.3H_{c2,T=0,{\varepsilon}=0}$.\cite{movie}  
In this simulation without applying external currents, 
new vortices are generated in the upper (positive $z$) surface 
continuously, and finally they move out to the lower 
(negative $z$) surface after moving to lower side. 
The vortices do not form stationary lattice permanently.
In our simulation, the local voltage 
($\propto \partial {\bf A}/\partial t$)\cite{MMachida} 
appears in the vortex.

The above-mentioned broken-centrosymmetric features of
the noncentrosymmetric superconductors appear,
when the magnetic field ${\bf H}$ is applied perpendicular to ${\bf n}$.
In the study of the vortex structure
when ${\bf H}$ is parallel to ${\bf n}$,\cite{Kaur,Hayashi}
the spin polarization ${\bf M}_{\rm s}$ appears
toward the radial direction around a vortex,
and the vortex structure remains to be cylindrical symmetric.

\section{SUMMARY}
We have shown how broken-centrosymmetric spatial structures appear
in the macroscopic scale of the noncentrosymmetric superconductors.
We investigate the surface structure in the Meissner state and
the vortex structure in the mixed state,
by numerical simulations of the TDGL equation.
When an applied field ${\bf H}$ is parpendicular to ${\bf n}$,
the broken-centrosymmetric spatial structures of the internal magnetic field
and the screening current are induced by the characteristic paramagnetic effect
due to the Rashba interaction.
Namely, the field penetration in the Meissner state becomes anisotropic
between $+{\bf n}$ and $-{\bf n}$-directions.
And the antisymmetric spin polarization ${\bf M}_{\rm s}$ around a vortex
breaks cylindrical symmetry of the vortex.
The paramagnetic supercurrent at a vortex center given by
the antisymmetric ${\bf M}_{\rm s}$ may induce
the flux flow without applying external currents.
We hope that these phenomena will be observed in future experiments,
as direct evidence of the noncentrosymmetric superconductors. 

\section*{ACKNOWLEDGEMENT}
We thank Hiroto Adachi for useful discussions.

\end{document}